CHAPTER 1

# Multi-Level Embedding Conformer Framework for Bengali Automatic Speech Recognition

**Md. Nazmus Sakib**

*Dhaka International University, Dept. of CSE, Dhaka 1212, Bangladesh.*

**Md. Maruf Bangabashi**

*Dhaka International University, Dept. of CSE, Dhaka 1212, Bangladesh.*

**Golam Mahmud**

*Dhaka International University, Dept. of CSE, Dhaka 1212, Bangladesh.*

**Umme Ara Mahinur Istia**

*Dhaka International University, Dept. of CSE, Dhaka 1212, Bangladesh.*

**Md. Jahidul Islam**

*Dhaka International University, Dept. of CSE, Dhaka 1212, Bangladesh.*

**Partha Sarker**

*Dhaka International University, Dept. of CSE, Dhaka 1212, Bangladesh.*

**Afra Yeamini Prity**

*Dhaka University of Engineering & Technology, Dept. of CSE, Gazipur, Bangladesh.*

CONTENTS

Bengali, spoken by over 300 million people, is a morphologically rich and low-resource language, posing challenges for automatic speech recognition (ASR). This research presents an end-to-end framework for Bengali ASR, building on a Conformer-CTC backbone with a multi-level embedding fusion mechanism that incorporates phoneme, syllable, and wordpiece representations. By enriching acoustic features with these linguistic embeddings, the model captures fine-grained phonetic cues and higher-level contextual patterns. The architecture employs early and late Conformer stages, with preprocessing steps including silence trimming, resampling, Log-Mel spectrogram extraction, and SpecAugment augmentation. The experimental results demonstrate the strong potential of the model, achieving a word error rate (WER) of 10.01% and a character error rate (CER) of 5.03%. These results demonstrate the effectiveness of combining multi-granular linguistic information with acoustic modeling, providing a scalable approach for low-resource ASR development.

## 1.1 INTRODUCTION

Automatic Speech Recognition (ASR) has become a cornerstone of modern human–computer interaction, powering a wide range of applications including voice assistants, transcription services, real-time translation, and accessibility tools for people with disabilities. Over the past decade, the advent of deep learning has significantly advanced ASR capabilities, particularly through Transformer-based architectures, which excel at modeling long-range dependencies and capturing complex temporal patterns in speech [1], [2]. These advancements have enabled highly accurate recognition in high-resource languages, facilitating the deployment of robust commercial and research-grade speech technologies. Despite these successes, progress in ASR has been uneven across languages. High-resource languages such as English, Mandarin, and Spanish benefit from extensive datasets, large-scale benchmarks, and mature research ecosystems [3]. In contrast, low-resource languages, including Bengali, suffer from limited annotated corpora, scarce linguistic resources, and complex linguistic structures that challenge conventional ASR approaches. This imbalance restricts access to effective speech technologies for millions of speakers and hampers efforts to build inclusive, global human–computer interaction systems [4]. Bengali, spoken by over 230 million people, ranks among the most widely used languages in the world [5]. However, its development for ASR remains challenging due to rich morphology, diverse phonetic inventory, and intricate syllable structures [6]. Tradi-tional ASR approaches for Bengali have often relied on purely acoustic modeling or simplistic subword units, which struggle to represent the nuanced interplay between pronunciation and linguistic patterns [7], [8]. Most existing works focus on single-level tokenization strategies—such as characters or wordpieces—thereby overlooking complementary linguistic information available at phoneme or syllable levels, which could enrich representation learning and enhance recognition accuracy.

To address these challenges, we propose an end-to-end Bengali ASR system that extends the Conformer-CTC backbone with a multi-level embedding fusion mechanism. Our approach integrates phoneme-, syllable-, and wordpiece-level embeddings into the encoder in a parallel, complementary manner. The system is structured into two stages: an early Conformer encoder extracts robust acoustic features, followed by a multi-level embedding stage where parallel Transformer-based networks encode linguistic information at multiple granularities. The fused representations are then refined by a late Conformer encoder and decoded using Connectionist Temporal Classification (CTC). This design allows the model to capture fine-grained phonetic cues while leveraging higher-level contextual patterns, offering a balanced and holistic representation suitable for morphologically rich languages like Bengali.
The contributions of this work are threefold.

- The design of an end-to-end Bengali ASR framework that integrates acoustic and multigranular linguistic representations.

- Empirical demonstration of the benefits of combining phoneme, syllable, and wordpiece embeddings for a low-resource, morphologically rich language.

- A scalable and adaptable architecture that can be extended to other underrepresented languages with complex linguistic structures.

- Integration of modern preprocessing techniques, including silence trimming, resampling, and log-Mel spectrogram extraction, to improve model robustness.

- Establishment of a scalable framework applicable to other low-resource languages facing similar ASR development challenges.

The remainder of this paper is organized as follows: Section 2 reviews related works in low-resource ASR and multigranular modeling; Section 3 details the proposed methodology; Section 4 describes the dataset and experimental setup; Section 5 reports and discusses the results; and Section 6 concludes with insights and future directions.

## 1.2 LITERATURE REVIEW

The advancement of Automatic Speech Recognition (ASR) has seen significant momentum with the evolution of deep learning models. Early ASR systems were predominantly based on statistical models such as Hidden Markov Models (HMMs) and Gaussian Mixture Models (GMMs), which required handcrafted features and extensive phonetic knowledge [9]. A Bangla speech recognition system was developed by Paul et al. [10], combining LPC-Cepstral feature extraction with a Multi-Layer Perceptron ANN, where SOM networks converted variable-length trajectories into fixed-length feature vectors and different architectures were compared for optimal performance. Hossain et al. [11] developed a Bangla speech recognition system for ten digits using MFCC features with a Back-Propagation Neural Network, achieving 96.33% recognition for known speakers and 92% for unknown speakers.

These traditional methods lacked scalability and struggled with high error rates in noisy or real-world environments. With the advent of deep learning, particularly the success of Recurrent Neural Networks (RNNs) and Long Short-Term Memory (LSTM) networks, ASR systems started achieving better performance due to their ability to model temporal dependencies [12]. An LSTM-based approach for Bengali word recognition was investigated by Nahid et al. [13], where MFCC features were extracted and processed through a deep LSTM with a softmax layer, achieving 13.2% word detection error rate and 28.7% phoneme detection error rate on the Bangla-Real-Number dataset. Mamun et al. [14] proposed an AI-enabled Bangla receptionist framework integrating multiple AI components, where the ASR model achieved a 9.01% WER and the system obtained over 75% user satisfaction. A Bangla STT system was developed by Saha [15] using a CNN-RNN architecture with CTC, trained on 215.53 hours of diverse speech data, yielding a gender- and speaker-independent model with reported WER results and implementation challenges discussed. An isolated Bangla word ASR system was developed by Rahman et al. [16] using MFCC features with DTW-based matching and SVM-RBF classification, achieving a recognition rate of 86.08%. Riswadkar et al. [17] developed a Bengali ASR system targeting low-resource settings, aiming to automate transcription in industries such as courts and local companies, reducing reliance on English-speaking stenographers. They customized the Wav2Vec2 model for semi-supervised training and evaluated performance using Word Error Rate (WER) and Character Error Rate (CER). Bengali ASR for the elderly domain, where domain-specific resources are scarce, was investigated by Paul et al. [10]. A non-invasive EEG-based BCI system for imagined Bengali speech recognition was developed by Hossain et al. [11], where a 14-channel headset and statistical feature extraction were employed, and a random forest achieved 84.28% (coarse) and 76.13% (fine) accuracy, surpassing prior studies.

A CNN-BiGRU network trained on general-domain data was adapted using transfer learning with just 5 hours of elderly speech, and the Character Error Rate (CER) was improved from 19.46% to 12.37%. The effectiveness of transfer learning for low-resource, domain-specific ASR development was thereby demonstrated. An Intellectual Bengali Speech Recognition System (IBSRS-DLT) was proposed by Deepa et al. [18] to address Bengali's morphological and dialectal challenges. However, these architectures had limitations in capturing long-range dependencies and required significant computational resources. Transformer-based models, introduced by Vaswani et al.[19] revolutionized sequence modeling by using self-attention mechanisms that can process input sequences in parallel and capture global dependencies effectively [20], [21]. This architecture inspired subsequent works in ASR, including models like Speech-Transformer and Conformer, which further enhanced performance by combining convolutional layers with self-attention mechanisms [22], [23]. The ability to capture contextual information across entire sequences has made transformers particularly effective for handling complex and variable-length speech inputs. Ahmed et al. [24] developed an automated system to convert handwritten Bangla characters into braille using a triple-stream hybrid model (ResNet, EfficientNet, and Vision Transformers) with a Character Quality Assessment Framework, achieving 95.27% validation accuracy and providing multilingual braille transcription with text-to-speech support. Ahnaf Mozib Samin [25] showed that using Byte Pair Encoding (BPE) with 500–1000 tokens for Bengali ASR improves out-of-distribution performance, outperforming character- and unigram-based tokenization, and reduces WER.

Despite advances in Bengali ASR, challenges remain, including limited datasets, morphological complexity, poor real-world robustness, and high computational demands. To address these gaps, this research develops a robust ASR system utilizing a Conformer-CTC backbone with multi-level embedding fusion of phoneme, syllable, and word piece representations. This approach improves accuracy across speakers and domains, effectively handles morphological variations, and enhances real-world performance efficiently.

## 1.3 METHODOLOGY

In our approach, special attention is given to handling the linguistic characteristics of Bengali, including conjunct consonants, diacritics, and morphological variations. The methodology integrates audio preprocessing, multi-level text representation, and deep sequence modeling to ensure accurate recognition across diverse speakers and dialects.

### 1.3.1 Proposed Framework

The proposed Bengali speech recognition framework, illustrated in Figure 1.1, follows a structured end-to-end pipeline consisting of preprocessing, multi-level modeling, and evaluation. The pipeline begins with raw audio input that undergoes preprocessing steps such as resampling, silence trimming, Log-Mel spectrogram extraction, SpecAugment augmentation, and normalization to generate robust acoustic features. These features are first processed by an early Conformer encoder, after which parallel embedding streams are introduced to capture different levels of linguistic information, namely phoneme, syllable, and wordpiece embeddings. A fusion mechanism combines these embeddings with the acoustic representation, which is then passed into a late Conformer encoder for deeper contextual modeling. Finally, a CTC decoder maps the processed features into text sequences. The model is trained on a Bengali speech corpus, and its performance is evaluated during inference using Word Error Rate (WER) and Character Error Rate (CER), ensuring reliable transcription quality.

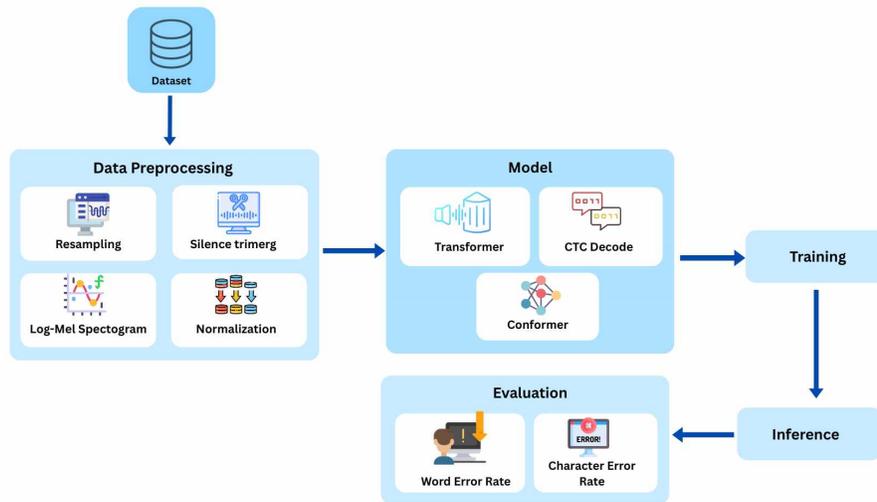

Figure 1.1 Proposed End-to-End Framework for Bengali Automatic Speech Recognition (ASR).

### 1.3.2 Dataset Collection

The OpenSLR Large Bengali ASR dataset was used to develop the Bengali Automatic Speech Recognition system. It contains high-quality WAV audio files and a TSV file with File ID, anonymized User ID, and transcriptions [26]. The dataset includes diverse speakers, accents, and recording conditions, enhancing model generalizability. With extensive linguistic variations and a large vocabulary, it is well-suited for training robust ASR models. Being open-source, it supports reproducibility and further research in Bengali language processing. Its size and variety help achieve better accuracy in speech-to-text applications. Table 1.1 presents the attributes of the dataset used in this study.

Table 1.1 Description of the dataset attributes.

| Attribute | Value |
|---|---|
| Number of utterances | 196k |
| Total Number of audio files | 2,04,905 FLAC audio files |
| Unique characters in transcribed text | 70 |
| Total unique words | 45,653 |
| Audio duration | 181.464 hours |
| Maximum text length | 30 |

### 1.3.3 Dataset Preprocessing

For audio preprocessing, the raw Bengali speech recordings were first resampled to a consistent rate to ensure uniformity. Leading and trailing silence were trimmed to remove unnecessary noise. Log-Mel spectrograms were extracted to represent the au-dio in a time-frequency format suitable for deep learning. SpecAugment was applied to augment the data by randomly masking time and frequency regions, improving model robustness. The features were then normalized to stabilize training and enhance convergence. For text preprocessing, the transcriptions were carefully cleaned and normalized to handle Bengali-specific challenges, such as conjunct consonants, diacritics, and spelling variations. Unwanted characters were removed, and the text was tokenized into phonemes, syllables, or wordpiece units, providing multi-level linguistic representations that account for the morphological richness of Bengali. The dataset was split into 80% for training, 10% for validation, and 10% for testing. Finally, these processed text sequences were aligned with the audio features for model training using the CTC loss.

### 1.3.4 Multi-level Embedding Enabled Convolution-Augmented Transformer

Our proposed Bengali Automatic Speech Recognition (ASR) framework is based on a multi-level embedding fusion architecture built upon the Conformer-CTC backbone. The system integrates phoneme-level, syllable-level, and wordpiece-level embeddings into the acoustic encoder to enhance linguistic representation. The overall pipeline is shown in Figure 1.2, while the internal embedding network is illustrated in Figure **??**.

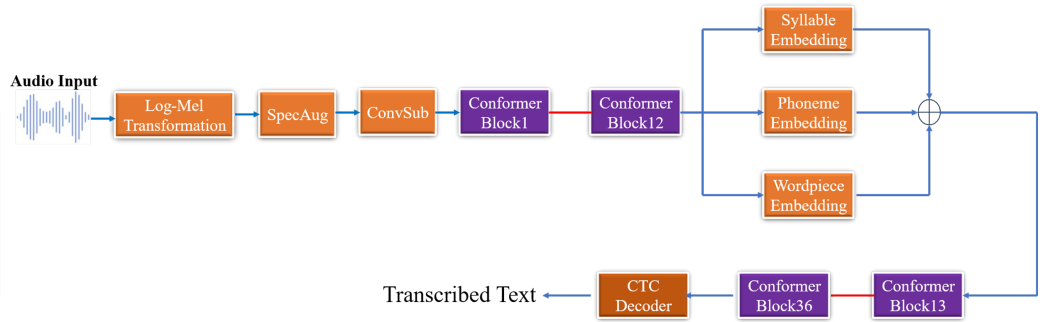

Figure 1.2 Architecture of the proposed Bengali speech recognition model.

Given an input audio waveform $x(t)$, the acoustic front-end computes log-Mel spectrogram features by applying a short-time Fourier transform (STFT) followed by Mel filterbank integration and logarithmic scaling. This transformation produces a feature matrix $F$:

$$F = \log\left(\text{Mel}(|\text{STFT}(x(t))|^2)\right) \tag{1.1}$$

where $F \in \mathbb{R}^{T \times D}$, with $T$ representing the number of frames and $D$ the Mel filterbank dimension. To improve generalization, SpecAugment is applied to mask portions of the spectrogram in both time and frequency domains, and a convolutional subsampling layer reduces the sequence length while enhancing local feature representation. The resulting features $F'$ are then passed through the early Conformer encoder, consisting of the first 12 Conformer blocks, which outputs:

$$H^{(E)} = \text{Conformer}_{1:12}(F') \tag{1.2}$$

This representation captures both local and global acoustic dependencies and serves as the input to the multi-level embedding modules. At this stage, we incorporated linguistic priors by branching $H^{(E)}$ into three parallel embedding networks corresponding to phoneme, syllable, and wordpiece levels. Each embedding module consists of a dense projection that maps the encoder output into an embedding dimension $d_e$, positional encoding to inject temporal ordering, a Transformer encoder for capturing contextual dependencies, and a final projection to produce embedding vectors. Formally, the embedding transformation can be expressed as:

$$E' = \text{Transformer}(H^{(E)}W_p + b_p + \text{PE})W_O + b_O \tag{1.3}$$

where $E'$ represents either the phoneme, syllable, or wordpiece embedding. These embeddings serve as complementary linguistic representations that enrich the acoustic features with information at multiple granularities. The outputs from the three branches are then merged with the acoustic features. In the simplest case, this fusion is realized by element-wise summation:

$$H^{(F)} = H^{(E)} + E_{\text{Phoneme}} + E_{\text{Syllable}} + E_{\text{Wordpiece}} \tag{1.4}$$

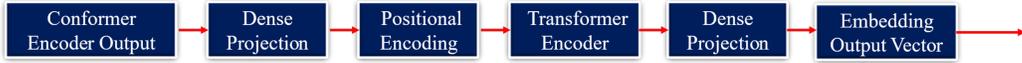

Figure 1.3 Internal structure of phoneme, syllable, and word embedding modules.

The fused representation $H^{(F)}$ is then fed into the late Conformer encoder, which consists of 24 Conformer blocks that further refine the joint acoustic-linguistic representation:

$$H^{(L)} = \text{Conformer}_{13:36}(H^{(F)}). \tag{1.5}$$

Finally, the processed sequence is projected into the vocabulary space and normalized via a softmax layer, yielding posterior probabilities over subword units. The model is trained with the Connectionist Temporal Classification (CTC) objective:

$$L_{\text{CTC}} = -\log \sum_{\pi \in \beta^{-1}(y)} \prod_{t=1}^{T'} Y_{t,\pi_t}, \tag{1.6}$$

where $Y$ denotes the predicted distribution, $\pi$ is a possible alignment path, and $\beta$ represents the collapsing function that maps frame-level alignments into the final transcription.

### 1.3.5 Hyperparameter Selection

Hyperparameters are predefined configurations that govern the learning process of a model but are not directly updated during training. They play a crucial role in controlling model complexity, convergence speed, and generalization ability [27]. In this study, a set of carefully chosen hyperparameters was used to optimize the performance of the proposed framework. Table 1.2 summarizes the ranges considered for each hyperparameter along with the selected values used in the final experiments. The model was trained with an appropriate number of epochs and batch size, while dropout was applied to reduce overfitting. The Adam optimizer, combined with gradient clipping and weight decay, ensured stable and efficient training. Different embedding types, phoneme, syllable, and wordpiece, were fused to capture richer linguistic features, and Log-Mel features were used as the acoustic input representation.

Table 1.2 Hyperparameters for Embedding-Enabled Conformer-based Bengali ASR

| Hyperparameter Name | Range | Selected Value |
|---|---|---|
| Epochs | 0–200 | 100 |
| Batch Size | 8–64 | 32 |
| Dropout Rate | 0.1–0.5 | 0.1 |
| Embedding Types | Phoneme, Syllable, Wordpiece | All three fused |
| Optimizer | Adam, AdamW, SGD | Adam |
| Learning Rate | 1e-5–1e-3 | 1e-4 |
| Loss Function | CTC, Cross-Entropy | CTC |
| Input Features | 40–128 Mel bands | 80-dim Log-Mel |
| Gradient Clipping | 1–10 | 5.0 |
| Weight Decay | 1e-6–1e-3 | 1e-5 |

### 1.3.6 Evaluation Metrics

Before introducing WER and CER, it is important to note that evaluation metrics are essential for assessing the performance of ASR systems. They measure how accurately the system transcribes speech compared to reference text, highlighting errors at different levels. Among these metrics, Word Error Rate (WER) evaluates word-level mistakes, while Character Error Rate (CER) focuses on errors at the character level, providing a more fine-grained assessment of transcription accuracy.

#### 1.3.6.1 Word Error Rate (WER)

The Word Error Rate (WER) is a widely used metric for evaluating the performance of Automatic Speech Recognition (ASR) systems, as it measures the percentage of incorrect words in the ASR output compared to a reference transcription. To calculate WER, the total number of words in the reference transcription ($N$) is first counted. Next, three types of word-level errors are identified in the ASR output: substitutions ($S$), which occur when a word in the output differs from the reference; insertions ($I$), which are extra words present in the output but not in the reference; and deletions ($D$), which represent words missing from the output that appear in the reference [28]. The total number of word errors ($E_w$) is then obtained as:

$$E_w = S + I + D \tag{1.7}$$

Finally, WER is calculated as the ratio of total word errors to the total number of words in the reference, expressed as a percentage:

$$WER = \frac{E_w}{N} \times 100 \tag{1.8}$$

This metric provides a standardized measure of ASR performance, indicating how closely the system's output matches the reference transcription.

### 1.3.6.2  Character Error Rate (CER)

Character Error Rate (CER) is another important evaluation metric for ASR systems, operating at the character level rather than the word level. It measures the percentage of incorrect characters in the ASR output compared to a reference transcription, taking into account substitutions, insertions, and deletions. To calculate CER, the total number of characters in the reference transcription ($N$) is first determined. Next, character-level errors are counted: substitutions ($S$), which are incorrect characters in the output; insertions ($I$), which are extra characters not present in the reference; and deletions ($D$), which are characters present in the reference but missing in the output [29]. The total number of character errors ($E_c$) is then calculated as:

$$E_c = S + I + D \tag{1.9}$$

Finally, CER is expressed as a percentage by dividing the total character errors by the total number of characters in the reference:

$$CER = \frac{E_c}{N} \times 100 \tag{1.10}$$

This metric provides a fine-grained assessment of ASR performance, ensuring precise transcription accuracy at the character level.

## 1.4  EXPERIMENTAL RESULTS

This section presents the performance of the proposed Bengali ASR system, evaluated using Word Error Rate (WER) and Character Error Rate (CER). The system was trained and tested on the OpenSLR Bengali corpus, following preprocessing steps such as silence trimming, resampling, and data augmentation. The experimental findings demonstrate that the proposed architecture achieves robust recognition performance, with low error rates that confirm its suitability for Bengali ASR in low-resource settings.

Figure 1.4 shows the WER and CER for the proposed Bengali ASR model. The system achieved its best performance with a CER of 5.03% and a WER of 10.01%, demonstrating robust and accurate transcription at both character and word levels.

Figure 1.5 illustrates the training and validation loss of the multi-level embedding Conformer model for Bengali ASR across epochs. The curve shows a steady decrease in loss, indicating effective learning and convergence of the model. The close alignment between training and validation loss suggests good generalization without overfitting.

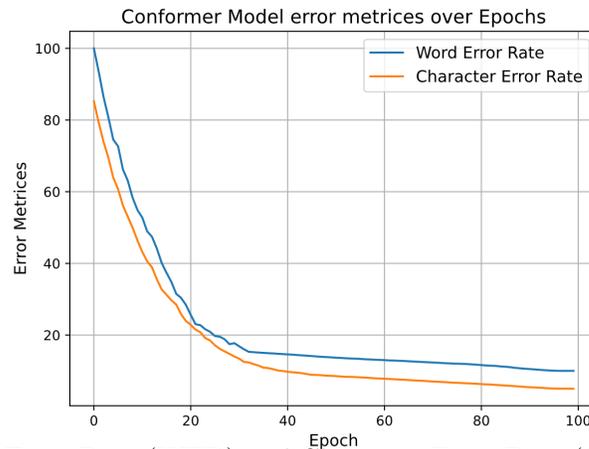

Figure 1.4 Word Error Rate (WER) and Character Error Rate (CER) of the proposed Bengali ASR mc

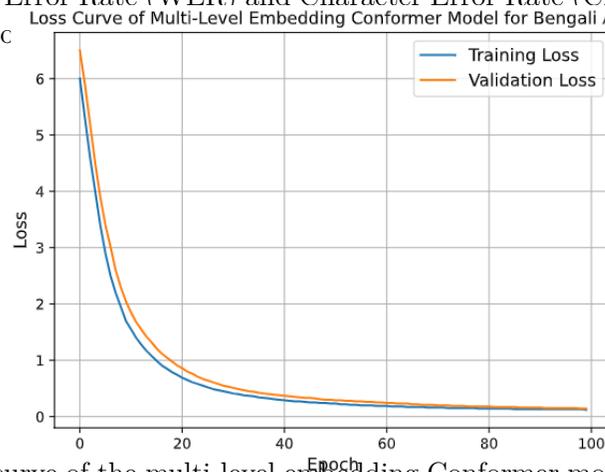

Figure 1.5 Loss curve of the multi-level embedding Conformer model for Bengali ASR over epochs.

## 1.5 COMPARATIVE PERFORMANCE ANALYSIS OF THE PROPOSED MODEL AGAINST RECENT STATE-OF-THE-ART APPROACHES

Table 1.3 presents a comparative performance analysis of the proposed Multi-Level Embedding Conformer model with other recent approaches for Bengali ASR. Prior works employing GRU, LSTM-RNN, and CNN-RNN reported relatively higher error rates, with WER values ranging from 21.18 to 54.19 on the OpenSLR Bengali dataset. Although Purkaystha et al. achieved a lower WER of 8.20, their evaluation was limited to isolated word-level predictions on the Bengali Broadcast Speech dataset, without complete sentence recognition. In contrast, our proposed model was evaluated on full sentence-level recognition using the OpenSLR Bengali ASR dataset and achieved a WER of 10.01 along with a CER of 5.03, demonstrating strong performance while ensuring robustness across both word- and character-level metrics.

Table 1.3 Comparative performance of the proposed Multi-Level Embedding Conformer model with recent SOTA Bengali ASR approaches

| Authors' Name | Dataset | Methodology | WER (%) | CER (%) |
| --- | --- | --- | --- | --- |
| Adhikary R et al. [30] | OpenSLR Bengali ASR | GRU | 54.19 | - |
| Nahim M et al.[13] | OpenSLR Bengali ASR | LSTM-RNN | 28.70 | 13.20 |
| B. Purkaystha et al.[6] | Bengali Broadcast Speech | LSTM | 8.20 | 3.00 |
| Sadeq S et al.[31] | OpenSLR Bengali ASR | RNN | 21.18 | 12.18 |
| Samin A. et al.[25] | OpenSLR Bengali ASR | RNN | 36.12 | 13.92 |
| Islam et al.[32] | Bengali News Corpus | CNN-RNN | 26.00 | - |
| Shahgir et al.[33] | Bangla Common Voice Speech | Wav2Vec2 | 25.24 | - |
| **Proposed Model** | OpenSLR Bengali ASR | **Multi-Level Embedding Conformer** | **10.01** | **5.03** |

## 1.6 CONCLUSION

This research presents a novel end-to-end Bengali speech recognition model based on a multilevel embedded conformer, specifically designed to tackle the challenges of accurately transcribing Bengali speech. The model was trained on the Large Bengali ASR Dataset from Open Speech and Language Resources, providing a rich and diverse dataset for robust learning. Experimental results demonstrate the model's strong potential, achieving a Word Error Rate (WER) of 10.01% and a Character Error Rate (CER) of 5.03%, indicating a significant improvement over existing approaches. These low error rates reflect the model's ability to effectively handle the complexities of the Bengali language, making it well-suited for applications such as transcription services, voice assistants, and accessibility tools.

Future work in Bengali speech recognition should aim to develop generalized models that can effectively manage the language's diverse dialects and regional variations, ensuring accuracy and accessibility across all Bengali-speaking areas. There is also a critical need to address real-world conditions by incorporating noisy, real-time data, such as phone call recordings. Improving model performance under these challenging scenarios will enhance the practical applications of Bengali ASR, including automated customer service and transcription systems.